\documentclass{cambridge6A}
\usepackage{graphics}
\usepackage[active]{srcltx}
\usepackage{amsfonts}
\usepackage{amsmath}
\usepackage{bm}
\input{epsf}
\usepackage{epsfig,here}

\newcommand{\uck}[1]{\o}

\newcommand{\ket}[1]{\mbox{$|#1\protect\rangle$}}

\newcommand{\bra}[1]{\mbox{$\protect\langle#1|$}}

\newcommand{\abs}[1]{\lvert#1\rvert}

\begin{document}

\chapter{Notes on Adiabatic Quantum Computers}

\begin{center} \large Boaz Tamir$^1$, Eliahu Cohen$^2$ \end{center}

 \begin{center} $^1$  Faculty of Interdisciplinary Studies, Bar-Ilan University, Ramat-Gan, Israel \\
                $^2$ H.H. Wills Physics Laboratory, University of Bristol, Tyndall Avenue, Bristol, UK \end{center}

We discuss in this chapter the basics of adiabatic computation, as well as some physical implementations. After a short introduction of the quantum circuit model, we describe quantum adiabatic computation, quantum annealing, and the strong relations between the three. We conclude with a brief presentation of the D-Wave computer and some future challenges.

\section{Introduction}

\begin{figure}[h]
\centering \includegraphics[height=6cm]{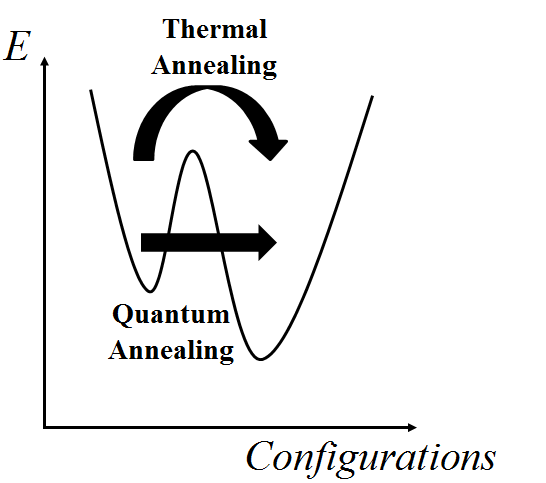}
\caption{Quantum annealing vs. thermal annealing in a graph of energy as a function of configuration space.}
\label{BTEC1}
\end{figure}

During the last two decades, a great deal of attention has focused on quantum computation following a sequence of results \cite{TCShor,TCGrover} suggesting that quantum computers are more powerful than
classical probabilistic computers. Following Shor's result \cite{TCShor}, that factoring and extraction
of discrete logarithms are both solvable on quantum computers in polynomial
time, it is natural to ask whether other hard (consuming exponential resources) problems
can be efficiently solved on quantum computers in polynomial time.
It was Feynman's idea \cite{TCFeynman} that quantum phenomena could not always be simulated
by classical computers, and whenever there are such simulations there is an exponential
growth in the required resources. Feynman also suggested the use of quantum computers and conjectured that quantum computers can be programmed to
simulate any local quantum system. Since then, a vast literature has been written,
addressing the theoretical and practical advantages of quantum computers, as well as some challenges in implementing them. In 1996 Lloyd
supported Feynman's claim and concluded \cite{TCLloyd}: ``The wide variety of atomic, molecular
and semiconductor quantum devices available suggests that quantum simulation
may soon be reality''. Just 3 years later, D-Wave systems were founded with the goal
of making practical quantum computers \cite{TCDwave}.
Indeed, quantum technology is maturing to the point where quantum devices,
such as quantum communication systems, quantum random number generators and
quantum simulators are built with capabilities exceeding classical computers.
Quantum annealers \cite{TCDas}, in particular, solve hard optimization problems by evolving a known initial configuration towards the ground state of a Hamiltonian encoding a given problem. Quantum annealing is an advanced alternative to classical simulated annealing \cite{TCLaarhoven}, an approach to solve optimization problems based on the observation that the problem's cost function can be viewed as the energy of a physical system, and that energy barriers can be crossed by thermal hopping. However, to escape local minima it can be advantageous to explore low energy configurations quantum mechanically
by exploiting superpositions and tunneling (see Fig. \ref{BTEC1}). Quantum annealing and adiabatic
quantum computation are algorithms based on this idea, and programmable
quantum annealers, such as the D-Wave computers, are their physical realization.
Quantum information processing offers dramatic speed-ups, yet is famously susceptible
to decoherence, the process whereby quantum superposition decays into
mutually exclusive classical alternatives, a mixed state, thus robbing quantum
computers of their power. For this reason, many researchers put in question the quantum features of the D-Wave computers \cite{TCBoixo1,TCBoixo2,TCAaronson}. In what follows we shall refer to the controversy concerning the quantum properties of the D-Wave computers.

In this short review work, we aim to present the crux of the subject matter. We shall focus on some fundamental results, leaving the small details outside. A strictly related, extensive work can be found in \cite{TCEPJST}.

\section{The Circuit Model}
\begin{figure}[h]
\centering \includegraphics[height=4cm]{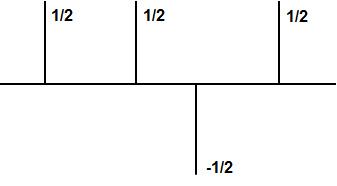}
\caption{The set of amplitudes following the first step of the Grover iteration in the 2 qubit case.}
\label{BTECa}
\end{figure}

As mentioned above, quantum computation was first suggested by Feynman as a way to overcome the
problem of simulating quantum phenomena on a classical computer \cite{TCFeynman}. Feynman
pointed out that a set of measurements on EPR entangled quantum particles could
not be simulated in principle by classical means. Moreover, even when one can use classical
computers to simulate quantum phenomena the growth in resources is exponential.
Therefore the natural way is to think of quantum computers. Soon after, Benioff \cite{TCBenioff} and Deutsch \cite{TCDeutsch1} presented a quantum version of a Turing machine (see also \cite{TCAlbert}).
However, the quantum Turing machine model was not practical. In 1989 Deutsch
suggested the idea of a quantum gate network computer \cite{TCDeutsch2}. He also provided a strong
argument showing that any finite dimensional unitary operator on a quantum state
could be simulated by a simple universal gate. Deutsch's universal gate was a 3 qubit gate, a variant of the known Toffoli gate for reversible classical computation. This universal gate approximates any
other quantum gate by using the well-known Kronecker \cite{TCKronecker} approximation. Deutsch
also presented the first known ``quantum algorithm'', later extended to the
Deutsch-Josza algorithm \cite{TCJozsa}. These algorithms can distinguish between a balanced
function and a constant one by using a small number of measurements. They showed
an exponential benefit over classical deterministic algorithms. In the scheme presented by Deutsch, quantum computers have no architecture and in that sense they resemble old,
one purpose, analogue computers. Following the work of Deutsch, two main
families of algorithms were introduced -- Grover's search and Shor's factoring. In 1996 Grover \cite{TCGrover}  presented a quantum search
algorithm for an element in an unsorted array. The Grover algorithm has a speedup
of a square root over the classical search algorithm (that is, if the size of the search space is $2^n$, then the Grover complexity is $\sqrt{2^n}$). Although such a speedup does
not cross a computational complexity class line ({\it i.e.} it does not turn a hard problem into a simple one), it shows a clear (and proven) gap between the
quantum and classical computational complexity. We can easily demonstrate the algorithm for the two qubit case. In general, the algorithm consists of $O(\sqrt{2^n})$ iterations, in the two qubit case one iteration is enough. Each such iteration consists of 2 substeps; the first marks the solution (without knowing its position, therefore using a black box) by a -1 phase, leaving all other elements unchanged, the second step is a reflection of each of the amplitudes over the (new) average of all amplitudes. In particular, for the 2 qubit case, following the first step, assuming the 3-rd element is the solution, we will get the amplitudes as in Fig. \ref{BTECa}.

\begin{figure}[h]
\centering\includegraphics[height=4cm]{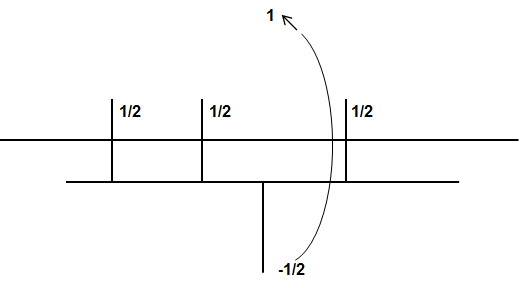}
\caption{The set of amplitudes following the second step of the Grover iteration in the 2 qubit case.}
\label{BTECb}
\end{figure}

Now the average of all the amplitudes is 1/4. Reflecting the 1/2 amplitudes over the 1/4 line brings them to 0, while reflecting the -1/2 amplitude over the same line bring it to 1 (see Fig. \ref{BTECb}). Hence one Grover iteration is enough.

Note that the algorithm is a ``black-box'' (or oracle algorithm), and as such can be
generalized and used to speed-up many classical algorithms \cite{TCBoyer}. Later, Grover also used
the above algorithm to present a scheme for the construction of any superposition in
a $2^n$ dimensional vector space using $\sqrt{2^n}$ steps at most \cite{TCGrover1}.
In 1994, Shor \cite{TCShor1} presented a polynomial algorithm for prime factorization (for a composite number $N$, having $logN=n$ digits, a non-trivial factor of $N$ could be found with $O((logN)^3)$ operations) and
discrete logarithms. So far, all known classical algorithms for the factorization or
discrete logarithm have exponential complexity. Therefore the reduction in complexity
seems exponential. However, we have no proof for the claim that the complexity
of such classical algorithms should be bounded from below by exponential function. This is a manifestation of the well-known ``P vs. NP'' problem (note however that factorization is not NP-complete).
Shor's algorithm used a Fourier transform module that can identify the order of a modular function.
This was the extension of a previous quantum algorithm by Simon \cite{TCSimon}. The
Fourier quantum module could serve also for phase estimation \cite{TCKitaev}, for order finding \cite{TCShor1,TCShor97},
and in general for the identification of other ``Hidden Subgroup'' of symmetries \cite{TCJozsa1}.

A severe drawback in quantum computation was and still is the problem of
decoherence \cite{TCNamiki}. It is hard to construct a stable superposition of even a small number of
qubits. It is even harder to apply unitary gates between the qubits. So far there are
several suggestions as to the way to construct a quantum computer. Clearly, it is
enough to construct the set of universal quantum gates (for the existence of such a set see \cite{TCDiVincenzo1}): a $XOR$ or what is known in quantum
computation as a $CNOT$ gate, and a one qubit rotation.

A major breakthrough came with the presentation of fault tolerant quantum gates \cite{TCShor2}. The basic ideas is the following: we first replace each qubit with a block of qubits, using some error correction code. The two physical states of the original qubit correspond to ``logical'' states of the block. Next, we write a ``logic'' gate between the blocks (a universal gate). Our aim is to control the propagation of noise, such that an error inside a block will not leak to the far away blocks, by employing fault tolerant gates to control it. These fault tolerant quantum gates and error correction quantum codes sustain the hope that one day a large scale quantum computer could indeed be realized.

Several criteria were suggested by DiVincenzo \cite{TCDiVincenzo2} for the physical possibility of the realization of quantum computers: \hfill\break \\
(1) One clearly needs a physical representation of qubits. \hfill\break
(2) The coherence time of the qubits should be large enough to allow the computation.\hfill\break
(3) There should be a physical mechanism realizing the unitary evolution of the
qubits. This mechanism must be controllable. \hfill\break
(4) Initial qubit states should be conveniently prepared. \hfill\break
(5) There should be a way performing a projective measurement of the final qubit
states. \hfill\break

In 2000, Farhi \cite{TCFarhi} described a new model of quantum computation based on the
quantum adiabatic theorem. It turned out that the quantum adiabatic model is
equivalent to the quantum gate network model of Deutsch \cite{TCAharonovd}. The adiabatic model is discussed in the next section.

Small scale quantum computers based on different kinds of physical qubits have
been implemented so far. To name just a few: single photon quantum computers \cite{TCChuang,TCKnill},
nuclear spins \cite{TCNuclear,TCVandersypen}, trapped ions \cite{TCCirac}, neutral atoms in optical lattices \cite{TCBrennen}, states of superconducting circuits \cite{TCMooij}, quantum dots \cite{TCImamog} and electrons' spin on Helium \cite{TCPlatzman}. We will focus on quantum adiabatic computation as the ``software'' of quantum computers, while for some ``hardware'' details we refer the reader to \cite{TCCohenTamir}.

\section{Adiabatic Computation}

Adiabatic quantum computation (AQC) is a scheme of quantum computation that is theoretically predicted to be more robust against noise than other methods \cite{TCChilds,TCPaz,TCSLloyd}. In this scheme a physical system, is initially prepared in its known lowest energy configuration, or ground state. The computation involves gradually deforming the system's Hamiltonian, very slowly, to assure that the system remains in its ground state throughout the evolution process. One designs the evolution of the Hamiltonian such that the ground state of the final Hamiltonian is the solution to the optimization problem. AQC is based on the adiabatic theorem stated by Born and Fock \cite{TCFock}:

{\it ``A physical system remains in its instantaneous eigenstate if a given perturbation is acting on it slowly enough and if there is a gap between the eigenvalues and the rest of the Hamiltonian spectrum.''}

The Hamiltonian is therefore time dependent $H=H(t)$. The initial Hamiltonian $H(0)=H_0$ and its lowest energy eigenvector should be easy to construct. We assume the final Hamiltonian $H_T$ is also easy to construct, however its ground state which is the solution to our optimization problem could be exponentially hard to find with a classical algorithm. The Hamiltonian of the AQC is therefore:
\begin{eqnarray}
H(t) =  \left(1- \frac{t}{\tau}\right)H_0 +  \frac{t}{\tau}H_T \equiv (1-s)H_0 +sH_T,
\end{eqnarray}
where $\tau$ is the adiabatic time scale, and $t$ goes from $0$ to $\tau$. The complexity of the adiabatic algorithm is manifested in the time it takes to evolve the computer from its initial to its final state. It can be shown \cite{TCFarhi}, that the adiabatic approximation is valid when the annealing time satisfies:
\begin{eqnarray} \label{ECBTadiabaticT}
\tau>> \frac {\max_{0\le s \le 1}[ \langle 1(s)| \frac{dH(s)}{ds} |0(s)\rangle]}{\min_{0\le s \le 1}[ \Delta_{10}(s)]^2},
\end{eqnarray}
where $|{i(s)}\rangle$ for $i=0,1$ are the ground and first excited states of $H(s)$, and $\Delta_{10}(s)$ is their energy difference. To understand the above lower bound on the time complexity we go back to basic principles \cite{TCMessiah}. Given a time dependent Hamiltonian we can write:

\begin{equation}
H \ket{\psi_n(t)}= E_n(t) \ket{\psi_n(t)},
\end{equation}

\noindent where  $\ket{\psi_n(t)}$  (resp. $E_n(t)$) is the n-th eigenvector (resp. eigenvalue). We can write a general solution to the Schr\"{o}dinger equation as:

\begin{equation}
\ket{\psi(t)} = \sum c_n(t)  \ket{\psi_n(t)} e^{i \theta_n(t)},
\end{equation}

\noindent where $\theta_n(t)$ is known as the dynamic phase. Using the Schr\"{o}dinger equation one can verify that:

\begin{equation}
\dot{c}_n(t) = -c_n(t) \bra{\psi_n(t)}{\dot{\psi}_n(t)}\rangle -
\sum_{m\ne n} c_m(t) \frac{ \bra{\psi_n(t)}\dot{H}\ket{\psi_m(t)}}{E_m - E_n} e^{i(\theta_m-\theta_n)}.
\end{equation}

\noindent Now, to ensure that the evolution of the n-th eigenstate remains in the n-th eigenstate, we have to reduce the amplitude:

\begin{equation}
\frac{ \bra{\psi_n(t)}\dot{H}\ket{\psi_m(t)}}{E_m - E_n}.
\end{equation}

\noindent This can be done by varying $H(t)$ very slowly with respect to $ {E_m - E_n}$. This is the origin of the argument for the complexity of the adiabatic computer. In fact, this also shows that one can preserve all eigenstates if the evolution is slow enough. In particular if this first gap decreases very rapidly  or exponentially  as a function of the number of variables in our problem, then we should exponentially slow down the evolution. A more accurate estimation of the complexity time using perturbation theory was suggested in \cite{TCJansen}. In \cite{TCYoung}, the minimal gap for the Exact Cover problem was studied using quantum Monte Carlo simulations, with the adiabatic computation implementation in mind. It turned out that the time complexity for the computation of the minimal gap is exponential.

In \cite{TCRigolin} an adiabatic theorem for degenerate states was discussed. A similar principle applies also there. If the evolution is slow enough then the transition between the eigenspaces, that is between different energy states has low amplitude. A necessary and sufficient conditions for the adiabaticity were computed.

AQC was first proposed by Farhi {\it et al.} \cite{TCFarhi}, where the 3-SAT problem was discussed.  It was also suggested in \cite{TCFarhi}, that one way to attack the time complexity problem is by looking for a tensor decomposition of the total Hamiltonian to smaller-dimensional Hamiltonians.

Can an adiabatic computer solve an NP-complete problem efficiently?  This question was not answered yet, however, the discussion in \cite{TCGutmann} is so far the closest as we can get to an answer.  There, the Exact Cover problem for a set of random instances was discussed. Each instance was constituted by several random iterations, where on each iteration a random clause was picked and added to the set of previously picked clauses, thereby reducing the number of satisfying assignments, down to only one. In this generating process the relation between the number of variables and the number of clauses is close to 1, this is believed to be the phase transition point between instances with several solutions (low number of clauses) and instances with no solutions (high number of clauses). Such instances are believed to be hard to solve on a classical computer. The quantum adiabatic algorithm was simulated on a classical computer up to $n=20$ (where $n$ is the number of variables). The time needed to get a success probability higher than some fixed value (1/8) was computed. It turn out that the time needed is quadratically related to $n$. Note that this quadratic relation is true only to low values of $n$. Moreover, it is not clear if the above randomly generated set of instances are really hard to compute on a classical computer, nevertheless the results in \cite{TCGutmann} are most challenging.

In \cite{TCFarhi2}, a possible illustration was given to the exponential time complexity of the AQC in solving 3-SAT problems. Given a $2^n$ dimensional space, a cost function was defined by taking the sum of a 3-local cost functions $h_3(z_i,z_j,z_k)$ on all sets of three variables. Such a cost function can be shown to be symmetric and a function of (only) the Hamming weight. The Hamiltonian therefore resembles a 3-SAT Hamiltonian. It turns out that the minimal eigenvector $\ket{\theta,s}$ can be parameterized by an angular parameter $\theta$. At some point $s^*$, there is a degeneracy, where $\ket{\theta_1,s}$ and $\ket{\theta_2,s}$ are two distance vectors with almost 0-eigenvalue.  At  $s<s^*$,  $\ket{\theta_1,s}$ is closer than $\ket{\theta_2,s}$to the 0-eigenvector and at $s> s^*$,  $\ket{\theta_2,s}$ is closer to the 0-eigenvector. For that to happen, the system, while crossing $s^*$  should tunnel through a barrier which might take an exponential time. One has to evolve the system slowly enough to allow the tunneling near $s^*$. Otherwise it will not stay in the global minimum. The time needed for such a tunneling can be computed by tunneling methods such as instantons. This emphasizes the importance of tunneling in quantum computing.

In \cite{TCvan_Dam} it was suggested that AQC resembles a local search: when given a problem where the global optimum lies in a narrow basin, while there is a close local minimum with a much larger basin, the AQC might need an exponential time to reach the global minimum. This is as if the AQC is stuck in the large basin's local minimum. Note however, that this is not the case, since an AQC always stays at the global minimum, and therefore the existence of a large basin local minimum should be translated to an exponential contraction of the first gap. To demonstrate this behaviour of the AQC, the authors suggested an artificial example where a global function was defined by the Hamming weight of its input configuration, being equal to this weight if it is lower than $\frac{1+\epsilon}{2}n$, and -1  (the global minimum) if the weight is higher.

In \cite{TCvan_Dam} it was also proved that one can simulate any AQC by a series of unitary gates. This is one direction of the proof of the equivalence between AQC and the circuit model, below we shall describe the second direction (from unitary gates to adiabatic evolution). The proof in \cite{TCvan_Dam} relies on the discretization of the continuous time dependent Hamiltonian into small intervals, where on each the instantaneous Hamiltonian is time independent.  It is easy to see that the norm difference between the continuous Hamiltonian and the discrete one is bounded above by an efficient function, and therefore the norm difference of the corresponding unitary gates (as proved in the paper) is also bounded by a similar bound. One has yet to show (as done in the paper) that each of the unitary gates could be efficiently constructed.

A major breakthrough was achieved when Aharonov {\it et al.} showed \cite{TCAharonovd} that adiabatic quantum computation is equivalent to the circuit model. Aharonov's theorem is based on the fact that for any circuit model algorithm we can produce a ``history vector'' describing the whole process of concatenated unitary gates on the initial vector. This history vector could be written as the ground state of a certain Hamiltonian, hence the circuit computation turns out to be an adiabatic one.

Here we shall present the main arguments given in \cite{TCAharonovd} since they have major significance for the whole subject.

Given a quantum circuit algorithm we have to show that there exists an adiabatic algorithm that can produce the same output vector with high probability. A circuit algorithm is a concatenation of $L$ unitary gates $U_l$ for $l=0$ to $L$. Suppose the output of $U_l$ is $\ket{\alpha_l}$, such that $U_l|\alpha_l\rangle = |\alpha_{l+1}\rangle$. The trick is to look at the history vector:

\begin{equation}
\ket{\eta} = \frac{1}{\sqrt{L+1}} \sum_{l=0}^L \ket{\alpha_l} \otimes\ket{l},
\end{equation}

\noindent where $\ket{l}$ denotes a clock vector in the l-th state, and the $\ket{l}$s  are orthogonal. In fact, such a vector was used by Kitaev in \cite{TCKitaev1} and was previously suggested by Feynman in \cite{TCFeynman1}. Consider the Hamiltonian $H_P$:

\begin{equation}
H_P = \frac{1}{2} \sum_{l=0}^L H_l
\end{equation}

\noindent where
\begin{equation}
H_l =
I\otimes \ket{l}\bra{l} - U_l\otimes \ket{l+1}\bra{l} - U_l^\dag\otimes\ket{l} \bra{l+1}+I\otimes\ket{l+1}\bra{l+1}.
\end{equation}

\noindent  By the definition of $H_l$ and the orthogonality of the $\ket{l}$s:
\begin{equation}
\begin{array}{lcl}
H_l ( \ket{\alpha_l}\otimes \ket{l}) =  \ket{\alpha_l}\otimes \ket{l} -  \ket{\alpha_{l+1}}\otimes \ket{l+1}\\
H_l ( \ket{\alpha_{l+1}}\otimes \ket{l+1} )= \ket{\alpha_{l+1}}\otimes \ket{l+1}- \ket{\alpha_l}\otimes \ket{l},
\end{array}
\end{equation}

\noindent and therefore $\ket{\eta}$ is the $0$ eigenvector of the Hamiltonian $H_P$ (one has to take care of the definitions of boundary conditions for $H_l$ and $\ket{\alpha_l}$ at $0$ and $L$). Note that $H_l$ is designed to verify that the history vector has the right concatenation at the $l-th$ point.

Now define

\begin{equation}
 \ket{\gamma_l}= \ket{\alpha_l}\otimes \ket{l}.
\end{equation}

\noindent We can then write:
\begin{equation}
 \frac{1}{2}H_l + \frac{1}{2}H_{l-1} = \ket{\gamma_l}\bra{\gamma_l} -
\frac{1}{2} \ket{\gamma_{l+1}}\bra{\gamma_{l+1}} - \frac{1}{2} \ket{\gamma_{l-1}}\bra{\gamma_{l-1}},
\end{equation}

\noindent and hence $H_P$ is a Toeplitz matrix on the space spanned by the $\ket{\gamma_l}$ for $l=0$ to $L$ :

\[ H_P = \left( \begin{array}{ccccccc}
.&.&.&.&.&.&.\\
-\frac{1}{2}&1&-\frac{1}{2}&.&.&.&.\\
.&-\frac{1}{2}&1&-\frac{1}{2}&.&.&.\\
.&.&-\frac{1}{2}&1&-\frac{1}{2}&.&.\\
.&.&.&-\frac{1}{2}&1&-\frac{1}{2}&.\\
.&.&.&.&-\frac{1}{2}&1&-\frac{1}{2}\\
.&.&.&.&.&.&. \end{array} \right). \]

\noindent It is left to compute the first gap of the $L+1$ dimensional matrix and show that it does not contract to $0$ faster than some polynomial function in $L$.

For small $s$ values, we can use the Gershgorin theorem \cite{TCBhatia}. For $s$ close to $1$ we have to use stochastic methods. Consider $G(s)=I-H_P(s)$ and note that $G(s)$ is primitive. Therefore, by the Perron-Frobenius lemma \cite{TCHorn}, it has a non-degenerate eigenvector for its highest eigenvalue $\mu(s)$  with positive entries $(\alpha_0,...,\alpha_L)$. This highest weight vector is also the ground state of $H_P(s)$.

Using $G(s)$ we can define a stochastic matrix $P=P(s)$:

\begin{equation}
P_{i,j}(s)= \frac{\alpha_j}{\mu(s) \alpha_i} G_{i,j}(s).
\end{equation}

\noindent The rest of the proof follows from the two arguments below:

a) The limiting distribution of $P(s)$ is  $(\frac{\alpha_0^2}{Z},...,\frac{\alpha_L^2}{Z})$, where $Z=\sum_i \alpha_i^2$ is a normalizing factor and the first gap of $P(s)$ is $\Delta H_P(s) /\mu(s)$.

Hence, the limiting distribution and the first gap of the above stochastic matrix is directly related to the ground state and the first gap of $H_P(s)$. Therefore we can use stochastic arguments to complete the proof. It is best to look at the conductance of the stochastic process. If $B$ is a subset in $\{0,...,L\}$, $P_{i,j}$ a stochastic process with limiting distribution $\pi$,  then the flow out of $B$ is defined by:
\begin{equation}
F(B) = \sum_{i \in B, j\not\in B} \pi_i P_{i,j},
\end{equation}

\noindent and the ``weight'' of $B$ is $\pi(B)=\sum_{i \in B} \pi_i$. The conductance of $P$ is then defined as:

\begin{equation}
\phi(P)=min_B \frac{ F(B)}{\pi (B)},
\end{equation}

\noindent where we minimize over all the subsets $B$ such that $\pi(B) < \frac{1}{2}$. This is the minimal normalized flow.


b) The conductance $\phi(P)(s)$ can be easily estimated to satisfy
\begin{equation}
\phi(P)(s) \geq \frac{1}{6L}.
\end{equation}

\noindent  This is shown by using the monotonicity of the coordinates of the ground state of $H_P(s)$. The theorem now follows from the known connection between the conductance and the first gap in stochastic processes.  Low conductance means a low first gap. In particular, the first gap is at least $\frac{1}{2} \phi(P)^2$ \cite{TCSinclair}. Hence we achieve the desired inverse polynomial condition for the first gap.

The computation ends where we measure the output of the adiabatic computer, {\it i.e.} the history vector. We can first measure the clock register, and if the clock points to $L$ then the vector register hold $\ket{\alpha_L}$.  To decrease the angle between the history vector and  $\ket{\alpha_L}$ (in fact, the embedding of $\ket{\alpha_L}$ in the history vector), one can add several identity gates at the end of the circuit network, then the same vector $\ket{\alpha_L}$ will appear in the last few vectors $\ket{\gamma_l}$.

Since this proof, several adiabatic protocols were suggested to solve, for instance, the problems of: Graph Isomorphism, Quantum Counting, Grover's search problem, the Deutsch-Jozsa problem and Simon's problem \cite{TCFarhi,TCvan_Dam,TCHen3,TCHen4}. However, in general, there is no direct and simple way to translate an algorithm in terms of the circuit model to an algorithm in terms of adiabatic computation, the proof in \cite{TCAharonovd} does not provide a simple way to go from one model to the other. This difficulty is mainly due to the fact that the exponentiation of a sum of Hamiltonians that do not commute is not the product of the exponentiation of each of the Hamiltonians. Therefore, in the circuit model we can present a simple set of universal gates, whereas for the adiabatic model it is much harder.

Albeit all the above difficulties, the equivalence between the models provides a new vantage point from which to tackle the central issues in quantum computation, namely designing new quantum algorithms and constructing fault tolerant quantum computers.

In \cite{TCChilds}, the robustness of AQC was discussed, using a master equation. A problem Hamiltonian solving an instance of a 3-bit Exact Cover problem was constructed using spin $\frac{1}{2}$ variables. There is an inherent robustness of the adiabatic evolution against dephasing in-between the eigenvectors since the computer stays in its ground state. Therefore, the problem is decoherence, due to the environment, into a higher eigenvector, or more generally, into a Gibbs state. In \cite{TCChilds}, a weak system-bath coupling was used. It was also assumed that the initial state of the system and bath is a tensor product of density matrices, and moreover, that the unitary evolution in the Master equation is governed by a time independent Hamiltonian. This last assumption is plausible since the evolution is very slow being adiabatic. The bath was assumed to be made of photons. An explicit Master equation was computed in the energy eigenstates. The results were rather expected. The success probability increased with computation time (being more adiabatic), also being larger for a higher gap problem. For high temperature the decoherence effect decreased the success probability. For very low temperatures and short computation times, the decoherence effect increased the probability by letting the system relax to the ground state from higher energy states, where it was placed due to the fast evolution. However, the results were non-conclusive being computed for very low dimensional instances.

In \cite{TCPonce} an Adiabatic Perturbation Theory (APT) was introduced. Such a theory expresses $\ket{\psi(s)}$ in terms of a power series in $v=\frac{1}{\tau}$ (where $\tau$ is the adiabatic time scale):

\begin{equation}
\ket{\psi(s)} = \sum_p v^p \ket{\psi^p(s)},
\end{equation}

\noindent where $\ket{\psi^p(s)}$ is the p-th perturbation term, and $\ket{\psi^0(s)}$ is the standard adiabatic term. For a very large $\tau$, the only term in the above sum will be the standard adiabatic one. We thus write $\ket{\psi(s)}$ in terms of the eigenvectors $\ket{n(s)}$ of $H(s)$:

\begin{equation}
\ket{\psi(s)} = \sum_{n,m} \sum_p v^p e^{\frac{-i}{v} \omega_m(s)} e^{i\gamma_m(s)} b_{n,m}^p(s) \ket{n(s)},
\end{equation}
\noindent where $\gamma_m$ is a dynamic phase and $\omega_m$ is a geometric Berry phase. The main result in \cite{TCPonce} is a recursion formula between $b_{n,m}^{p+1}(s)$, $\dot{b}_{n,m}^p(s)$, and $b_{k,m}^p(s)$ for all $k$. This enables the computation of the $p+1$ perturbation term using the $p$ terms.

 Such a perturbation theory is a major step in computing adiabatic sufficient and necessary conditions.

\subsection{Simulated annealing and adiabatic computation}

There is a deep analytic relation between simulated annealing and quantum adiabatic processes. Below we describe these relations \cite{TCOrtiz2,TCMorita1}, also known as the classical to quantum mapping. Suppose we are given a set of Boolean variables $\sigma_i =\pm 1$ defined on an $n$ dimensional lattice. Let $E$ be an energy (cost) function defined on the $2^n$ dimensional configuration space $\overline{\sigma} = (\sigma_1,..,\sigma_n)$, for example

\begin{equation}
E(\overline{\sigma}) = \sum_{i,j} J_{i,j} \sigma_i \sigma_j,
\end{equation}

\noindent where $J_{i,j}$ are coupling coefficients. One can define a stochastic process $S_\beta$ (for $\beta = \frac{1}{k T}$, an inverse temperature) such that its limiting distribution $\pi_\beta$ satisfies:

\begin{equation}
\pi_\beta(\overline{\sigma}) = Z_\beta^{-1} e^{-\beta E(\overline{\sigma})}
\end{equation}
\begin{equation}
Z_\beta = \sum_{all~\overline{\sigma}} e^{-\beta E(\overline{\sigma})}.
\end{equation}

\noindent If $\beta$ is big enough, then sampling from $\pi_\beta$ will result in a lowest energy configuration with high probability. Now for the quantum analogue; For each $\overline{\sigma}$ we can construct a corresponding quantum pure state $\ket{\overline{\sigma}}$, defined in the tensor space of Pauli z-spinors. Set $S_\beta$ to be the stochastic matrix satisfying the Detailed Balance Condition:
\begin{equation}
S_\beta({\overline{\sigma}_i}|{\overline{\sigma}_j}) \pi_\beta(\overline{\sigma}_j)=
S_\beta({\overline{\sigma}_j}|{\overline{\sigma}_i}) \pi_\beta(\overline{\sigma}_i).
\end{equation}

We can now define a Hamiltonian by setting its matrix coefficients to:

\begin{equation} \bra{\overline{\sigma}_i}H_\beta \ket{\overline{\sigma}_j} = \delta_{i,j} - \sqrt{ S_\beta({\overline{\sigma}_i}|{\overline{\sigma}_j}) S_\beta({\overline{\sigma}_j}|{\overline{\sigma}_i})}.
\end{equation}

\noindent Define now the state

\begin{equation}
\ket{\psi_\beta} =  \frac{1}{\sqrt{Z_\beta}}\sum_{all~\overline{\sigma}} e^{-\beta E(\overline{\sigma})/2}  \ket{\overline{\sigma}}.
\end{equation}

\noindent Then $\ket{\psi_\beta}$ corresponds to the Gibbs state for $E$ and $\beta$. It turns out  that  $\ket{\psi_\beta}$ is the unique ground state of $H_\beta$ \cite{TCOrtiz1}. In such a correspondence one can also show that a transverse field component such as $\sum_j \sigma_j^x$, is only natural in this correspondence between the stochastic transition matrix and the Hamiltonian. In fact, the ground state of $I- \frac{1}{n}\sum_j \sigma_j^x$  which is the sum of all states in the computational z-basis with equal amplitudes, corresponds to the completely mixed state at infinite temperature.

We can now let $T=T(t)$ be a function of time. The question is how to choose the correct pace to decrease $T$. In \cite{TCOrtiz2} \cite{TCMorita1} the evolution rate was computed to guarantee that the whole process is adiabatic. The correct rate was found to be (no faster than)

\begin{equation}
T(t) \sim \frac{n}{log (t)},
\end{equation}

\noindent which is very close to the known simulated annealing rate. Therefore, it seems that this correspondence between stochastic matrices and Hamiltonians embed simulated annealing into the set of quantum annealing, as if a special path of the quantum annealing process (in fact an adiabatic one) can produce a classical simulated annealing process.

In \cite{TCKnill2} a spectral gap amplification method was suggested for classical annealing algorithm. This is done by quantum simulation of the classical annealing process using a version of the Grover algorithm suggested in \cite{TCAmbainis}. While solving the problem of Element Distinctness a new version of the Grover algorithm was suggested. The algorithm alternate between two type of transformations, the first marks the target state (the ground state of the final Hamiltonian) using a -1 phase and leaves any orthogonal state untouched (see the discussion in the previous chapter), while the second transformation leaves the initial (start) state untouched while multiplying any of its perpendicular vectors by the same phase, this corresponds to the reflection transformation in the original Grover model. It was shown in \cite{TCAmbainis} that such alternate iterations take the start state close to the target state. In  \cite{TCKnill2} this method was used to artificially amplify the spectral gap.

\subsection{Different paths from an initial to a final eigenstate}

An alternative way to the standard continuous path evolution of the AQC, was suggested in \cite{TCBoixo123}. Using the quantum Zeno effect \cite{TCZeno}, one can discretely evolve the initial state to the final state or very close to it. In the quantum Zeno effect, we need $\frac{L^2}{1-p}$ steps to cross the angular distance $L$ with fidelity $p$. We use a discrete set of Hamiltonians $H(l)$, $0\le l \le L$, and we guarantee that the evolution by each of the Hamiltonians is close to the Zeno projection into $\ket{\psi_l}$, a non degenerate eigenvector of $H(l)$. Each Hamiltonian is evolved for a time $t$, which is randomly chosen from some distribution. Let $R^t_l$ denote the corresponding evolution operator of $H(l)$ for a time $t$. We define a projective measurement operator by:

\begin{equation}
M_l (\rho) = P_l \rho P_l + (I-P_l) \rho (I-P_l),
\end{equation}

\noindent where $P_l = \ket{\psi(l)}\bra{\psi(l)}$.  One can now estimate the difference between the projective measurement operator and the evolution operators:

\begin{equation}
||(M_l - R_l^t) (\rho) ||_{tr} \leq sup_{\omega_j} \Phi({\omega_j}),
\end{equation}
\noindent where $\omega_j$ are the energy difference to the other eigenvalues of $H(l)$, and $\Phi$ is the characteristic function of the distribution of $t$. We can now estimate the expectation value of the distribution of $t$,  {\it i.e.} the average time we need to evolve any of the Hamiltonians, in terms of $\Phi$ (which lies in the frequency domain). With some knowledge on the random variable $t$ we can write

\begin{equation}
\langle t \rangle \gg \frac{1}{min_s \Delta (s)} = \frac{1}{\Delta}.
\end{equation}

\noindent Using the above condition on $\langle t \rangle$, we can guarantee that the random evolution operators are close to the projective measurement operators, it then follows that the complexity of the whole process
is:

\begin{equation}
O\left( \frac{(L)^2 log(L/(1-p))}{(1-p) \Delta}\right).
\end{equation}




\subsection{Imaginary time and simulations}

An analysis of imaginary time was presented in \cite{TCMorita1}. It was found that the asymptotic behavior of the imaginary time quantum annealing (IT-QA) is the same as the real time quantum annealing (RT-QA), also the error of the IT-QA is no larger than the error of RT-QA. However, the importance of the use of imaginary time lies in the fact that this new algorithm can be simulated on a classical computer and could be considered as a form of a quantum Monte Carlo method.

This is the essence of Quantum MC techniques;  we use imaginary time in the Schr\"{o}dinger equation, turning the quantum equation into a classical one. Some of the common methods are the Variational \cite{TCMcMillan}, Diffusion \cite{TCGrimm}, Auxiliary Field MC \cite{TCCeperley}, Path Integral \cite{TCBarker}, Gaussian \cite{TCCorney}, and Stochastic Green Function \cite{TCRousseau}




\subsection{Complexity and universality}

A few words on complexity class theory and universality are in order. The quantum parallel of the classical NP complexity class is known as QMA \cite{TCWat}. It is the class of all languages $L$ that can be probabilistically verified by a quantum verifier in polynomial time. In particular, let $B$ be the Hilbert space of some qubit, $V$ a polynomial time quantum verifier and $p$ a polynom. We say that $L$ is in QMA if the following two conditions are satisfied:

a) If $x\in L$ then there exists a proof $\pi$ in $B^{p(\abs{x})}$ such that:
\[ pr(V(x,\pi) = 1) > \frac{2}{3} \]

b) If $x \not\in L$ then for all $\pi$ in $B^{p(\abs{x})}$;
\[ pr(V(x,\pi) = 1) \leq \frac{1}{3} \]

The connection between the above definition of QMA and our discussion goes through the definition of the $k$-Local Hamiltonain Problem. Given a $k$-local Hamiltonian, that is, a Hamiltonian that acts on only $k$ qubits, suppose it is promised that the ground state of the Hamiltonian is either below $a$ or above $b$ where $a<b \in [0,1]$ and $1/(b-a) = O(n^c)$ for some constant $c$. Then one has to distinguish between the two cases.

A $k$-local Hamiltonian can be thought of as a set of local constrains on the set of $n$ qubits. Therefore, the problem of $k$-local Hamiltonian resembles the MAX-k-SAT problem. Kitaev showed that the $log(n)$-local Hamiltonian problem is in QMA \cite{TCKitaev1}, where $\abs{x} = n$. Moreover, the 5-local Hamiltonian is QMA-complete. This fact could be interpreted as a map between the language of computational complexity theory and the language of condensed matter physics, hence its importance. Some refinments of Kitaev's theory quickly followed; in \cite{TCKempe1} it was proven that the 3-local Hamiltonian problem is also QMA-complete, and in \cite{TCKempe2} it was shown that the 2-local Hamiltonian problem is QMA-complete.

It was later proven by Biamonte \cite{TCBia1} that the 2-local Hamiltonian problem is QMA-complete even when restricted only to real valued Hamiltonians (that is, represented by real matrices). Biamonte also showed that one can approximate the ground state of such Hamiltonian by a set of simple and realizable 2-local Hamiltonians; two such universal sets were introduced; in particular it was shown that:

a) The 2-local ZZXX Hamiltonians are QMA-complete, where:

\[H_{ZZXX} = \sum h_i \sigma_i^z +\sum \Delta_i \sigma_i^x +\sum J_{i,j} \sigma_i^z \sigma_j^z + \sum K_{i,j} \sigma_i^x \sigma_j^x \]

b) The 2-local ZX Hamiltonians are QMA complete, where:

\[H_{ZX} = \sum h_i\sigma_i^z +\sum \Delta_i \sigma_i^x +\sum_{i<j} J_{i,j} \sigma_i^z \sigma_j^x + \sum_{i<j} K_{i,j} \sigma_i^x \sigma_j^z \]

\noindent  In that context, a general scheme representing the ground states of  $k$-local Hamiltonians using 2-local Hamiltonians was presented in \cite{TCBia2}. This resembles the use of simple Karnaugh maps in reduction of variables.

\subsection{Additional methods}

In  \cite{TCOrtiz1} a ``partly adiabatic-partly diabatic'' process was suggested. Suppose the two lowest eigenvalues are separated from the rest by a polynomially decreasing gap, while the first gap decreases exponentially with the size of the problem. One can compute the probability of jumping to the second eigenvector if the evolution is too fast (polynomial) near the (normalized) time $s$ where the gap is minimal. In a similar way, one can compute the probability to go back to the ground state later on. This could be easy in case the problem is symmetric with respect to $s$. Such an example for the random glues tree problem was discussed in \cite{TCOrtiz1}. \hfill\break

A few final remarks before we go on to discuss quantum annealing. The adiabatic model has several setbacks. The most important is the lack of a guaranteed fault tolerant method. In the circuit model one can control the amount of noise passed on to neighboring qubits (see for instance the review in \cite{TCNielsen}). We can concatenate circuits where on each we control the amount of noise. This can not be done in the adiabatic case. This is connected to the fact that the adiabatic computer model has no universal subset of computers. However, the adiabatic computer is robust against several types of noise as discussed above.

One last remark, we desire to do the adiabatic evolution in zero temperature. In practice, if $kT$ is much smaller than the gap then the adiabatic evolution will overcome thermal noise. In general, if $kT$ is larger than the gap it might be useful to describe the evolution of the Hamiltonian within the context of the Master equation \cite{TCLidar-Master}.

\section{Quantum Annealing}

Quantum annealing was suggested as an improvement of the simulated annealing
technique which suffers a severe setback in cases where the system is ``non-ergodic''
({\it e.g.} systems described by the spin glass model). In such cases, configurations of $n$
spins corresponding to minimum of the cost function could be separated by $O(n)$
sized barriers \cite{TCMoore}, so that at any finite temperature thermal fluctuations takes practically infinite time to relax the system to the global minimum.

There are clear similarities between simulated and quantum annealing. In both methods, one
has to strictly control the relevant parameters and change them slowly to tune the
strengths of thermal or quantum fluctuations. In addition, the main idea behind
both classical and quantum annealing is to keep the system close to its instantaneous ground state. Quantum annealing excels in tunneling through narrow (possibly cuspidal) barriers. Classical simulated annealing schedules might still have an advantage where the barrier is wide and shallow.

The basic scheme is as follows. First the computational problem has to be mapped to a corresponding
physical problem, where the cost function is represented by some Hamiltonian $H_0$ of the Ising form:
\begin{eqnarray} \label{BTECIsing}
H_0=-\sum_{i<j}J_{ij}\sigma_i^z\sigma_j^z-\sum_ih_i\sigma_i^z,
\end{eqnarray}
where $J_{ij}$ denotes the coupling strength between spins $i$ and $j$, and $h_i$ describes the
magnetic field at site $i$. Then a suitably chosen non-commuting quantum tunneling
Hamiltonian $H_1$ is to be added,
\[ H_1 = \sum_i \Delta_i\sigma_i^x \]
\noindent where $\Delta_i$ denotes the interaction strength with the ``tunneling'' term, so that the total Hamiltonian takes the form of:
\begin{eqnarray}
H=H_0-\Gamma(t)\sum_i\Delta_i\sigma_i^x\doteq H_0+H_1(t),
\end{eqnarray}
where $\Gamma(t)$ describes $H_1$'s
time dependence. One can then solve the time dependent Schr\"{o}dinger equation for the
wave-function $\ket{\psi(t)}$:
\begin{eqnarray}
i\hbar\frac{\partial}{\partial t} \ket{\psi}=[H_0+H_1(t)]\ket{\psi}.
\end{eqnarray}

The solution approximately describes a tunneling dynamics of the system between
different eigenstates of $H_0$. Like thermal fluctuations in classical simulated annealing,
the quantum fluctuations owing to $H_1(t)$ help the system to escape from the local
``trapped'' states. Eventually $H_1(t)\rightarrow 0$ and the system settles in one of the
eigenstates of $H_0$; hopefully the ground state. This serves as a quantum analog of cooling
the system. The introduction of such a quantum tunneling is supposed to make the high
(but very narrow) barriers transparent to the system, and it can make transitions to
different configurations trapped between such barriers, in course of annealing. In other
words, it is expected that application of a quantum tunneling term will make the free
energy landscape ergodic (see the review in \cite{TCDasCh}), and the system will consequently be able to visit any configuration with finite probability. Finally the quantum tunneling term is tuned to zero to get back the Ising Hamiltonian. It may be noted that the success of quantum annealing is directly connected to the replica symmetry restoration in quantum spin
glass due to tunneling through barriers.

The fact that one can use the quantum tunneling effect produced by a transverse field to help evolve the system into its ground state was first suggested in \cite{TCRay} in 1989. It was initially contested by the work in \cite{TCAltshuler} on the ground that the Anderson localization will not alow it.

\subsection{The relation between simulated annealing, quantum annealing, and adiabatic computation}

This relation between simulated and quantum annealing was discussed {\it e.g.} in \cite{TCRose}. Assume first a classical system at a temperature $T$ is examined. We would like to minimize its free (Helmholtz) energy. If $p(x)$ is the probability of the state $x$, then eventually the distribution $p=p(x)$ will minimize the free energy:
\[ F_T(p) = E_p(H)- T S(p), \]
\noindent where $E_p(H)$  is the expected value of the energy function,
\[ E_p(H) = \sum_x p(x)  H(x), \]
\noindent and $S(p)$ is the entropy:
\[ S(p) = - \sum_x p(x) log [p(x)]. \]
\noindent There are two considerations in reaching the minimum value (distribution) of $F_T(p)$. First, the entropy should be maximized, this occurs when the probability of each state is the same. Second, $E_p(H)$ should be minimized. The shape of $E_p(H)$ is the landscape of the energy or cost function. When $T$ is big enough then the entropy is the dominant part in the expression for the free energy, and hence the equipartition of the probabilities is most important. In other words, if the temperature is high enough we can reach any point in the landscape of the cost function. If $T$ is reduced we discover the minima of the energy function, we can then hopefully descend into the global minimum. This is what we do in the simulated annealing algorithm, as well as in the physical process of annealing.  For the quantum case we look at a similar free energy function, this time over density matrices:
\[ F_T(\rho) = tr \rho (H)- T S(\rho). \]
\noindent When the temperature is $0$ we are left with  $tr \rho (H)$. We can write $H$ as a sum of two terms, ``kinetic'' and ``potential'' :
\[ H = K + V. \]
\noindent Now we can let $K$ play the same role as the entropy above. If, for example, we let
\[ K = -\sum_i \sigma_i^x, \]
\noindent then the minimal eigenvector of $K$ is an equal sum of all elements in the $z$ computational basis. We can also add a coefficient $\Gamma$ to control the amplitude of $K$. This is all done in the $0$ temperature case and therefore
 \[ F_T(\rho) = tr \rho (V) + \Gamma tr \rho(K). \]
\noindent This resembles a classical annealing process- we can start with high $\Gamma$ and slowly reduce into $\Gamma = 0$, where we hope to end in a global minimum.

To sum up, we have a phase space of two variables $T$ and $\Gamma$, $(T,\Gamma)$. Now $(T,0)$ is the classical annealing path, and $(0,\Gamma)$ is the quantum annealing path. Can we find different paths where both coefficients are non-zero with a low computation time complexity?

The quantum annealing, as evident from the discussion above, does not have to be adiabatic. Being adiabatic restricts the evolution time to be slow enough.

In \cite{TCFinnila} a tunneling effect was discussed for a double well cost function, where one of the wells has a lower minimum. The authors used the Diffusion Monte Carlo method, where an imaginary time is used in the Schr\"{o}dinger equation, turning it into a classical diffusion function. Thereafter one can use random walk agents to simulate the diffusion. This is a simple way to compute the ground state. The results were rather surprising and the tunneling effect was clearly shown;  an initial function located in the higher well was able to leap over the barrier into the lower energy state.

What are the differences between simulated and quantum annealing from the point of view of computational complexity theory? A convergence criterion was proved in \cite{TCMorita} which is similar to the well-known one in simulated annealing \cite{TCGeman}. If we let $\Gamma (t)= t^{-\gamma/ n}$, (where $\gamma$ is some positive constant) we are guaranteed to get a solution. This, however, could take an infinite amount of time. If we stop the relaxation at some final time $t_f$ where the ``temperature'' $\Gamma (t)$ is small, $\Gamma (t)= \delta\ll 1$ then it is enough to wait until $t_f=e^{-n ln(\delta)/2\gamma}$.  Compare this to the relaxation time $t_f$ for the simulated annealing protocol, there $T(t) \sim \frac{n}{k log(t)}$ (see the discussion in the previous section), and if $T(t_f) =\delta$, then $t_f= e^{\frac{n}{\delta k}}$.  Clearly, for very small $\delta$, {\it i.e.} $\frac{1}{\delta} >> -ln(\delta)$, the quantum annealing scheme will be better than its simulated annealing counterpart. This is true in general, but could be hard to utilize, since both relaxation times are exponential.

In fact, for some specific problems the advantage of quantum annealing over simulated annealing is much more clear. In \cite{TCKadowaki} it was tested on a toy model of 8 qubits with a transverse Ising field. The authors showed that quantum annealing leads to the ground state with much larger probability than the classical scheme, where the same annealing schedule is used. In \cite{TCMartonak} path-integral Monte-Carlo quantum annealing showed better results for the Traveling Salesman Problem for 1002 cities. Here the algorithm was stopped after various number of steps and the results were compared to a simulated annealing algorithm. QA was shown to anneal more efficiently, and to decrease the solution residual error at a much steeper rate than SA. The authors in \cite{TCFarhi2} constructs an example where the width between local minima is small and therefore the tunneling effect is strong. The simulated annealing counterpart of the example shows an exponential computational complexity.

Recent results suggest that for first order phase transitions the adiabatic algorithm has exponential time complexity. For second order phase transition the adiabatic algorithm has polynomial complexity. It was also suggested that by adding an annealing term one can solve the first order transition problems \cite{TCJorg,TCSeoane}.

Brooke {\it et al.} \cite{TCBrooke} applied the above model to a disordered ferromagnet. Their aim was to find the
ground state for the ferromagnet with a certain proportion of randomly inserted antiferromagnetic
bonds. Cooling it down to 30 mK and varying a transverse magnetic field, they were able to compare simulated and quantum annealing, concluding eventually that their experiment directly demonstrates the power of a quantum tunneling term in the Hamiltonian.

Another evidence for the existence of tunneling effects was presented in \cite{TCBoixo2}, where the D-Wave computer was tested on a family of randomly generated Ising problems. It showed a clear distinction between easy and hard problems. In the hard cases where the success probability was low, the Hamming distance between the final vector and the ground state was high, whereas in the easy cases (high success probability) the Hamming distance was low. This could suggests a tunneling effect. The hard cases are those that the evolution is too fast for them, hence the computer get stuck at an excited eigenvector, however the fact that these are exactly the states with high Hamming distance means that the tunneling was avoided there. For if the Hamming distance is $d$, then to cross that distance by tunneling is $\Gamma^d$ (exponentially) hard (this is clear if we look at the transverse field as a perturbation).

There is also a deep connection between the number of free qubits in the ground and first excited eigenstates and the first gap \cite{TCBoixo2}. It is easy to see that the transfer field breaks the degeneracy of a free qubit. Thus, if the first excited state has more free qubits than the ground state, the splitting of the energy states by the transverse field are such that the minimal gap reduces. This does not happen in classical simulated annealing. Therefore, such problems could be harder for quantum simulated annealing than for classical annealing.

\section{The D-Wave Computer}

On May 2011, D-Wave Systems Inc. announced ``D-Wave One'', as ``the
world's first commercially available quantum computer''. The D-Wave one contained 128 qubits. It provoked an
immediate controversy about its true properties. Recently
Google has purchased ``D-Wave Two'' containing 512 qubits.

\begin{figure}[h]
\centering\includegraphics[height=4cm]{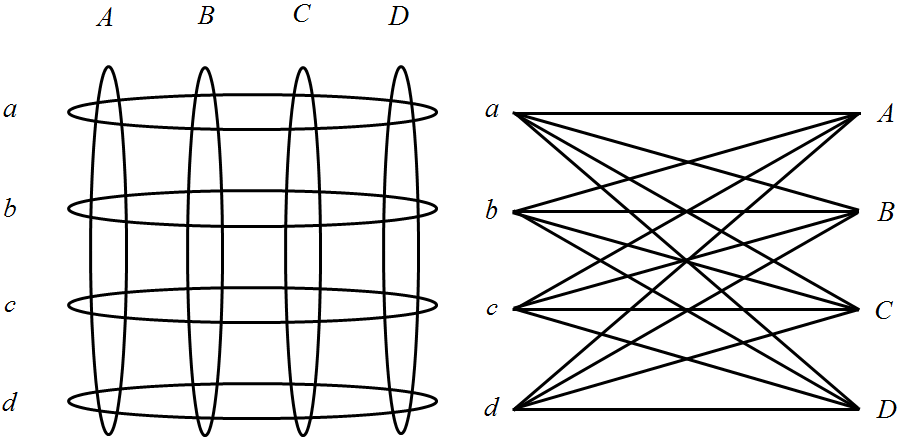}
\caption{8 qubit cell.}
\label{BTEC2}
\end{figure}

The D-Wave computers utilize flux qubits of the PCQ type \cite{TCWendin}. A set of 8 qubits are inter-coupled into a cell. In Fig. $\ref{BTEC2}$ qubit $a$ is coupled to qubits A,B,C and D. Similarly, qubit A is coupled to qubits a,b,c and d. All 8 qubits and their interconnections can be described by the graph in Fig. \ref{BTEC2}. In D-Wave Two, 64 such cells constitute a two dimensional grid. Each cell is connected to its neighboring cells. The whole 512 qubits therefore implement a graph known as the Chimera graph $C_n$.

\begin{figure}[h]
\centering \includegraphics[height=4cm]{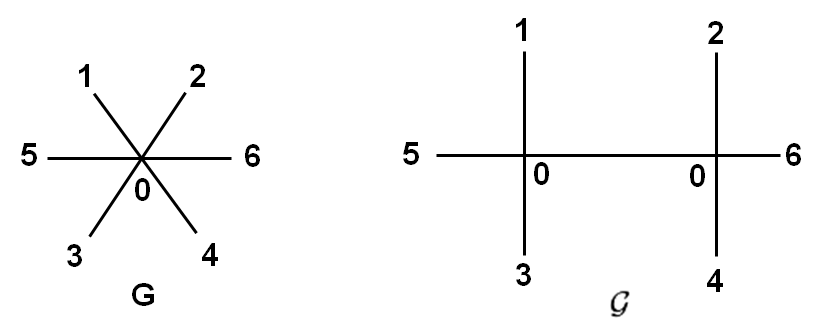}
\caption{The vertex $0$ in G is mapped into two such vertices to satisfy the physical requirements of $C_n$, that is, no more than 4 couplings to each physical qubit.}
\label{BTEC3}
\end{figure}

Since $C_n$ is not a complete graph it is not clear how to implement a general graph $G$ into the hardware. One should distinguish between logical qubits and physical ones. For example, if $C_n$ could only be connected to 4 of its neighbors, assume $G$ has a vertex $v_i$ with degree higher than 4 , then to implement $G$ inside such $C_n$ type machine we first need to map $v_i$  into  a subtree of several such vertices. We will use the leaves of the tree to connect to other vertices (see Fig. \ref{BTEC3}). On the hardware we will get a graph $\mathcal{G}$ of the physical qubits. The original graph $G$ is called the Minor of the expanded graph $\mathcal{G}$. Such a simple embedding demands that $|G|^2 \sim |\mathcal{G}|$, however there are more efficient embeddings (the task of finding such an embedding is in itself a hard computational problem, see also \cite{TCChoi}).

The qubits (spins) are coupled together using programmable elements which provide an energy term that is continuously tunable between ferromagnetic and anti-ferromagnetic coupling, this allows spins to favor alignment or anti-alignment, respectively. The behavior of this system can be described with an Ising model Hamiltonian, similar to the one in Eq. \ref{BTECIsing} above.

The computer is cooled to 20 mK by a dilution refrigerator, the pressure is set to $10^{-10}$ atmospheric pressure, and the computer is shielded magnetically to $5 \cdot 10^{-5}$ of earth magnetic field. The D-Wave computer has the size of a small chamber, however its core 512 qubits board is much smaller.

The ``Orion'' prototype of the D-Wave contained only 16 qubits. Some of the problems it solved were pattern matching, Seating Arrangement and a Sodoku puzzle \cite{TCSodoku}. The ``D-Wave One'' (incorporating 128 qubits) was used by a team of Harvard University researchers to present some results of the largest protein folding problem solved to date \cite{TCOrtiz}. The current ``D-Wave Two'' consists of 512 qubits  (not all active), and thus enables the solution of much more complex problems like network optimization and radiotherapy optimization which were demonstrated by the company \cite{TCDwave}.

In \cite{TCMcGeoch} it was claimed that on Quadratic Unconstrained Binary Optimization problems the D-Wave hardware returns results faster than the best known IBM CPLEX Optimizer, by a factor of about $3600$ for a problem of size $N=439$. This was contested by CPLEX developers in \cite{TCJeanFrancois}. Several similar claims were recently made by the D-Wave group, but so far for each such claim there is a contesting one questioning the results.

The D-Wave computer is introduced as a quantum annealer, or as an adiabatic computer having a programmable transverse field for tunneling. In general, it is hard to guarantee that the time evolution will meet the requirements of the adiabatic theorem, and indeed it turns out that the D-Wave computer is manifesting a regime which is in-between adiabatic and thermic. The excitations to higher eigenvectors in the course of evolution is expected to be followed by a later relaxation into a ground state. It was claimed by the D-Wave group that such a regime could even improve the probability to get the correct result \cite{TCDickson}.

\subsection{Is the D-Wave a Quantum Computer? The Future of Quantum Annealers}
\label{ECBTsec:6}

As a new apparatus it is only natural to ask how can we be sure this machine is indeed a quantum computer. For two parties having two computers we can play the Clauser, Horne, Shimony, Holt (CHSH) game, and if the success probability is higher than classical (75\%) we can assume the parties are using some quantum procedure with entanglements (see \cite{TCB.Reichardt}), however, for one computer the question is still open.  In what follows we present several criteria which we believe are important for the identification of the D-Wave (or any other computer) as a quantum computer. Some of the criteria can be attributed to the D-Wave computer, a few cannot, and the rest are still controversial.

\subsection{Universality}
\label{ECBTsec:6.1}

For the adiabatic model there is no natural set of universal gates generating the whole theoretic spectrum of the model. This we can call ``inner universality''. Since there is no such inner universality the computer resembles the old analog ``one purpose'' computers \cite{TCDas}.  Nevertheless we can ask what kind of problems can the D-Wave computer solve? Currently, only particular optimization problems are considered. However, as was shown in \cite{TCFarhi}, adiabatic computation can solve the 3-SAT problem (in exponential time) and therefore, in principle, any NP-hard problem could be solved. The (polynomial) equivalence of adiabatic computation and quantum circuit computation \cite{TCAharonovd} suggests that, theoretically, the adiabatic method can be generalized towards solving any problem that can be solved by the circuit model. Indeed, several protocols for solving specific problems other than optimization were suggested: Graph Isomorphism, Quantum Counting, Grover's search problem, the Deutsch-Jozsa problem and Simon's problem
\cite{TCFarhi,TCvan_Dam,TCHen3,TCHen4}. However, in practice, there is no direct and simple way to translate an algorithm in terms of the circuit model to an algorithm in terms of adiabatic computation (then again, see the method suggested by Lloyd in \cite{TCSLloyd}). Therefore, in the circuit model we can present a simple set of universal gates, whereas for the adiabatic model it is much harder.

\subsection{Coherence time of the SQUIDs}
\label{ECBTsec:6.2}

The coherence time of the qubit should be larger than the time needed for the algorithm to compute. This is far from being achieved in the D-Wave computer. The coherence time of the SQUID is about 10 ns while the annealing time needed is 5-15 $\mu$s \cite{TCBoixo2}. Indeed, how can one achieve quantum computation when the annealing time (depending on the first energy gap) of the computer is about 3 orders of magnitude longer than the predicted single-qubit coherence time? This seems to force a thermodynamic regime on the computer. The main reason for using such flux qubits is that they are relatively simple to manufacture, using common methods of lithography. A set of such SQUIDs, their coupling apparatus and measurement gates are concatenated as on a printed circuit board. Hence, we are still in need for the qubit -``transistor'', that is, a simple apparatus presenting a behavior of a two-state system, that can maintain coherence for a long time (in comparison to the operation time), and can be read off, coupled, and easily manipulated.

It seems that the D-Wave computer operates under a semi-adiabatic-semi-thermodynamic protocol. Near the anti-crossing the evolution might be too fast for an adiabatic computer. Therefore, the eigenvectors are excited to a higher energy states, later to be relaxed to the $0$ eigenstate again.  It was suggested by the D-Wave group that such relaxations could even help in the adiabatic evolution into the ground state \cite{TCDickson}.

Recently, a promising progress was achieved regarding the decoherence time of flux qubits \cite{TCStern}.

Another promising direction for increasing coherence time, is the research on anyons \cite{TCanyons}. These quasiparticles are topologically protected against decoherence and hence might be very valuable as the building blocks of a quantum annealer.

\subsection{Scalability}
\label{ECBTsec:6.3}

How many qubits can a D-Wave have? D-Wave computers have made a major leap when incorporating the largest number of qubits ever seen on a single device. The question now is of scalability. It is possible that the complexity to construct such a computer with all its inter-couplings, grows itself exponentially. This will mean that the possible gain in algorithmic complexity is paid out in building a coherent circuit (see also \cite{TCKalay}). This question is deeply connected to the lack of fault tolerant gate theory for adiabatic computation. A scalable architecture of adiabatic computing was suggested in \cite{TCKaminsky}, by translating NP hard problems to the Max Independent Set problem. For that problem a highly robust Hamiltonian was suggested. A more fundamental research in this issue should be done in the context of the master equation (see also \cite{TCLidar-Master}).

\subsection{Speed-up}
\label{ECBTsec:6.4}

Are the D-Wave computers faster than other computers running different optimization algorithms? For which problems? Right now, it seems that the answer to the first question is ``sometimes'' and it is not clear enough what is the answer to the second problem \cite{TCLidar}. In our opinion this is the most important indicator because of its practical significance, but currently it is a problematic issue \cite{TCLidar,TCKatzgraber}. In 2013 it was indeed admitted by the D-Wave group \cite{TCd-wave1} that different software packages running on a single core desktop computer can solve those same problems as fast or faster than D-Wave's computers (at least 12,000 times faster than the D-Wave for Quadratic Assignment problems, and between 1 and 50 times faster for Quadratic Unconstrained Binary).

In \cite{TCDefDet} the question of defining and detecting quantum speedup was discussed. It was implicitly suggested that one should think of a new way to define computational complexity, at least for cases where the instances of the computational problem are randomly generated. In the following we describe the general idea.

Consider the data in \cite{TCLidar,TCnewLidar}, where 1000 different spin-glass instances (randomly picked) where investigated. Each instance was run 1000 times and the success probability $s$ for finding the correct solution was computed. The parameter $s$ could also describe the ``hardness'' of the problem.

Suppose one annealing run takes $t_a$ time and has a success probability $s$. Therefore the total success probability of finding the solution at least once in $R$ runs is $ p = 1- (1-s)^R$. Set now $p=0.99$ and write $R=R(s)$. Let $T_{DW}(N,s)$ be the time complexity of the D-Wave computer wired to a problem of size $\sqrt{N}$ (see the discussion above on the Minor problem), and hardness $s$, and let $T_C(N,s)$ be the corresponding classical time complexity. Clearly $T_{DW}(N,s)$ would be proportional to $R t_a$. One way to define a speedup would be to look at the quotient of quantiles:
\begin{equation}
\frac{T_{DW}(N,s)|_{s\leq s_0}}{T_{C}(N,s)|_{s\leq s_0}}.
\end{equation}
\noindent This means that we average both complexities on a large set of instances (indexed by their hardness $s$) and only then compute the quotient. This suggests a new way to look at computational complexity theory, as a quotient of integrals or averages. Another way to define the speedup would be to look at the quantile of the quotient:
\begin{equation}
 \frac{T_{DW}(N,s)}{T_{C}(N,s)}\bigg |_{s\leq s_0}
\end{equation}

\noindent which compares the complexity of both computers on the same instance and only then as a function of hardness $s$. Both methods presented inconclusive results of speedup, although the second showed a small advantage for using the D-Wave computer when $N$ is large \cite{TCDefDet}.

\subsection{``Quantumness''}
\label{ECBTsec:6.5}

Since there is no clear evidence for a speedup, there is a possibility of comparing the behavior of the D-Wave computer to other models of computation with respect to a large family of computational problems. Consider again the data in  \cite{TCLidar,TCnewLidar}. A histogram describing the number of instances for each success probability $s$ was presented. The D-Wave histogram was found to be strongly correlated with quantum annealers (in fact simulated quantum annealers) rather than classical annealers. Both the D-Wave and the quantum annealer had a bimodal histogram, a large set of problems which are very easy to solve (high success probability) and a large set that are hard to solve (low success probability). The classical simulated annealer had a normal distribution type of histogram with respect to success probability (hardness to solve). This was considered as a proof for the quantumness of the D-Wave machine.

Note that by the above success probability distributions, a problem that is hard for one computer can be easy for the other, while the distribution for the whole ``hardness'' may look the same. This, in itself, questions the interpretation given to the above results.

These conclusions were also criticized by J. Smolin and others \cite{TCShin,TCSmolin1}. It was claimed that the difference between the histograms could be explained out on several grounds. Simulated annealing algorithms start from different initial points each time, while the adiabatic algorithms start from the same point and evolve almost the same each time. Hence, different adiabatic trials naturally show more resemblance. This implies that time scales for the simulated annealing algorithm and for the adiabatic algorithm could not be compared as such. It would be of interest to increase the number of trials given to the SA. This way, one could probably find a good correlation between the simulated annealing and the D-Wave.

In fact, in \cite{TCShin,TCSmolin1} a classical simulated annealing model was presented on a set of 2-dimensional vectors, a compass $O(2)$ model (an SD model). Indeed the model showed a bimodal behavior with respect to success probabilities \cite{TCSmolin1}. Being a classical model, this questions the quantum interpretation of the above results regarding the D-Wave computer.

In \cite{TCBoixo2} the correlation between the success probabilities of solving the same problem instance on any two computers in the set (SQA, DW, SD, SA) was computed. Note that this time each single instance was tested on two computers. High correlation between the DW and SQA (simulated quantum annealer) was shown. However, in \cite{TCShin} similar correlations (even slightly better) were presented between the classical $O(2)$ model and the D-Wave, suggesting a classical behavior of the D-Wave.

In \cite{TCBoixo1} the D-Wave One was tested on an artificial problem of a set of 8 spins: 4 core spin and 4 ancillae. The ground space for the particular wiring presented was highly degenerate and had two components, one was a cluster of 16 states, the second was a singular separate state. In the simulated annealing case writing a Lindblad set of equations shows that the separated state is enhanced {\it i.e.} the probability to end in that state is higher than the average probability to end in one of the other ground states. In short this could be explained by the fact that the separate state is close (in the Hamming distance) to a high number of (first) exited states, while any of the other ground states are close to a lower number of exited states. Since the evolution is thermic the computer easily jumps into exited states and relaxes back with higher probability into the separated ground state. The overall result is an enhancement of the separated state. As for the quantum adiabatic computer, in the midst of the evolution the separated state is no longer a ground state, due to the transverse field added. The state ``joins'' the ground space only at the end of the evolution, however, there the transverse field is too low to swap it with one of the other ground state. The overall result is an attenuation of the separated state. Hence we get a clear distinction between the two models. We can therefore use this toy problem as a test for the quantumness of the computer, and indeed  the D-Wave One showed the expected adiabatic behavior. In response, it was shown in \cite{TCShin} that the $O(2)$ classical model of the same problem exhibits a distribution of $0$ eigenvectors similar to the one presented by the adiabatic computer, although it is a classical computer.

As another proof for quantumness, the response of the computer to a change in the properties of the flux qubits, was suggested in \cite{TCJohnson}. For each qubit the thermal fluctuations are proportional to $e^{-\epsilon(U)/kT}$, where $\epsilon(U)$  is the barrier height (see the description of PCQ SQUIDs in \cite{TCWendin}). If we increase $\epsilon(U)$  the thermal fluctuations gradually stop until they freeze out at some freezing time $t_0^c$, such that  $\epsilon(U)(t_0^c)= kT$. Similarly, the tunneling effects also freeze out when $\epsilon(U)$  is increased above some value  $\epsilon(U) (t_0^q)$.  We expect the freezing time $t_0^c$ of the thermal fluctuation to be linearly dependent on temperature, whereas the tunneling freezing time $t_0^q$ to be independent of temperature.  The authors thereby apparently proved the existence of a tunneling quantum effect.

\subsection{Does the computer exhibit entanglement?}
\label{ECBTsec:6.6}


Recently Smirnov and Amin \cite{TCSmirnov} introduced a theorem that connects the magnetic susceptibility of the adiabatic Hamiltonian with the existence of entangled states. Suppose we define the susceptibility of a qubit $i$ to be:

\begin{equation}
\chi_i^\lambda = \frac{\partial \langle \sigma_i^z\rangle}{\partial \lambda}, \
\end{equation}
where $\lambda$ controls the evolution of the Hamiltonian (such as a time parameter). Suppose $\chi_i^\lambda $ and  $\chi_j^\lambda$ are both non-zero, $J_{ij} \neq 0$, and suppose the evolution is slow enough to reside on the ground state (there exists an anti-crossing), then the theorem states that at some  far point in the evolution process the eigenstate is entangled.

In \cite{TCLanting} the above theorem was tested on the D-Wave computer, for two toy models: a two qubit circuit and a cell of 8 qubits. Both the ground and the first exited states turn out to be entangled vectors. The measurement was done using qubit tunneling spectroscopy \cite{TCBerkley}.

Note however, that the sole existence of entanglement in the process of computation does not guarantee the quantum properties we need from a quantum computer, therefore this criterion is weak.

\subsection{Open Questions and possible future routes}

We believe that at this stage, where much work is still undergoing, we can only conclude this chapter with a few open questions. The D-Wave group has definitely made a great progress in the field, both on theoretical and practical aspects. However, the D-Wave computer is now at the apex of a controversy.

To sum up, we wish to raise several questions and ideas concerning future research. \hfill\break
(1) Choosing the hardware or the gates of a quantum computer, there are two main factors to be considered: the coherence time and the operation time. There should be a high relation between the two. The superconducting flux circuits of the D-Wave are far from being the best in that point. For coherence and operation time scales of other qubits see \cite{TCNielsen,TCLadd}, for achieving long coherence time (0.1 ms) in superconducting qubits see \cite{TCRigetti}. \\
(2) The benefits of the flux qubits of the D-Wave are clear: they are easy to build using known techniques of lithography, the flux qubits are easy to couple, etc. However, with respect to other computational properties they are only moderate \cite{TCLadd}. \\
(3) If the D-Wave computer has quantum properties, and also thermic properties, then the best way to analyze its behaviour is by Markovian Master equations (see also \cite{TCLidar-Master}). \\
(4) In \cite{TCKatzgraber} it was suggested that the glassy Chimeras of the D-Wave might not be the right architecture for testing quantum annealing. It seems that its energy landscape near zero temperature is too simple and does not have significant barriers to tunnel through. This attenuates the properties we want to use in the quantum computation. \\
(5) It could be that the Chimera graph of the D-Wave makes the embedding of graphs into the computer hard. Different wiring of the computer could make it easy to test other problems \cite{TCChoi}. \\
(6) One can test the quantumness of the D-Wave computer by testing its performance on a specific problem having a large (known) computational complexity gap between its classical annealing and quantum adiabatic versions. A simple version of such a test function was suggested by \cite{TCFarhi2}. It was demonstrated there that the time complexity of a classical simulated annealing computer solving such a problem is exponential due to the height of some spikes, while an adiabatic computer could easily tunnel (that is, in polynomial time) through the spikes if these are narrow enough. \\
(7) It could be useful to simulate other quantum informational tasks besides optimization, and even to test the D-Wave with hard fundamental tasks such as area law behavior etc. \cite{TCWolf}.\\


{\bf Acknowledgements}
\\ E.C. was supported by ERC AdG NLST.

\end{document}